\documentclass[prr,twocolumn,superscriptaddress]{revtex4-2}
\usepackage{amsmath,amssymb,dsfont,mathtools}
\usepackage[utf8]{inputenc}
\usepackage[T1]{fontenc}
\usepackage[english]{babel}

\begin{document}
\title{Periodic systems have new classes of synchronization stability}
\author{Sajad \surname{Jafari}}\email{Corresponding author: sajadjafari83@gmail.com}
    \affiliation{Health Technology Research Institute, Amirkabir University of Technology, Tehran, 1591634311, Iran}
    \affiliation{Department of Biomedical Engineering, Amirkabir University of Technology, Tehran, 1591634311, Iran}
\author{Atiyeh \surname{Bayani}}
    \affiliation{Department of Biomedical Engineering, Amirkabir University of Technology, Tehran, 1591634311, Iran}
\author{Fatemeh \surname{Parastesh}}
\author{Karthikeyan \surname{Rajagopal}}
    \affiliation{Centre for Nonlinear Systems, Chennai Institute of Technology, Chennai, 600069, Tamil Nadu, India}
\author{Charo I. \surname{del Genio}}
    \affiliation{Institute of Smart Agriculture for Safe and Functional Foods and Supplements, Trakia University, Stara Zagora 6000, Bulgaria}
\author{Ludovico \surname{Minati}}\email{Corresponding author: lminati@uestc.edu.cn}
    \affiliation{School of Life Science and Technology, University of Electronic Science and Technology of China, 611731 Chengdu, China}
    \affiliation{Center for Mind/Brain Sciences (CIMeC), University of Trento, 38123 Trento, Italy}
\author{Stefano \surname{Boccaletti}}
    \affiliation{Sino-Europe Complex Science Center, North University of China, Shanxi, Taiyuan 030051, China}
    \affiliation{Institute of Interdisciplinary Intelligent Science, Ningbo Univ. of Technology, Ningbo, China}
    \affiliation{CNR - Institute of Complex Systems, I-50019 Sesto Fiorentino, Italy}

\date{\today}

\begin{abstract}
The Master Stability Function is a robust and useful tool for determining
the conditions of synchronization stability in a network of coupled systems.
While a comprehensive classification exists in the case in which the nodes
are chaotic dynamical systems, its application to periodic systems has been less
explored. By studying several well-known periodic systems, we establish a
comprehensive framework to understand and classify their properties of
synchronizability. This allows us to define five distinct classes of synchronization
stability, including some that are unique to periodic systems. Specifically,
in periodic systems, the Master Stability Function vanishes at the origin,
and it can therefore display behavioral classes that are not achievable in
chaotic systems, where it starts, instead, at a strictly positive value.
Moreover, our results challenge the widely-held belief that periodic systems
are easily put in a stable synchronous state, showing, instead, the common
occurrence of a lower threshold for synchronization stability.
\end{abstract}

\maketitle

Over the last couple of decades, the most successful structural
paradigm in the study of complex systems has been that of networks,
in which discrete elements called \emph{nodes} or \emph{vertices}
interact across connections called \emph{links} or \emph{edges}~\cite{Alb02,Boc06,Boc14,Boc23}.
Of particular relevance to real-world applications is the case
where the nodes are dynamical systems, coupled to each other if
they share an edge. However, proper frameworks
and techniques are required to operationally define robustness
and resilience of networks leading to optimal performance~\cite{Art24}.
In dynamical networks, a vast array of phenomena
can occur, driven by the collective organization of the individual
dynamical systems. A significant one is the emergence
of a synchronized state, in which a number of elements that can
extend to the entire network eventually converge to the same trajectory
in phase space~\cite{Pik01,Boc02,Boc18}. The study of synchronized
states holds a special importance across fields, as it has found
notable applications such as in modelling the functioning of neurons
and the brain, and in investigating and optimizing the operation
of power grids~\cite{Var01,Fra10,Nis15,Tot20,Hov20,Sch22,Xu23,Pal24}.
As a result, strong efforts have been directed towards the study
of the different forms under which synchronized states appear and
of the effects that factors such as network structure and coupling
configuration have on their properties~\cite{Lag00,Pik00,Bar02,Nis03,Bel04,Per04,Cha05,Hwa05,Mot05,Yao09,Li13,Mas13,Pec15,Aro18,Rak19,Rak20,Sug21,Kor21,Fra22,Naz23,Bay23,Maj22}.

A related question, which has generated a large body of work,
is how to assess the stability of a synchronized state. A powerful
tool to address this problem is the method known as the Master
Stability Function~(MSF)~\cite{Pec98}. The method estimates the
stability of the synchronous solution to the dynamics by estimating
the largest Lyapunov exponent after the system is perturbed in
directions within the subspace transverse to the synchronization
manifold. This allows one to evaluate the synchronization stability
from the sign of the exponent, so that the trajectory of the
perturbed system will converge back onto the synchronized state
only if the largest Lyapunov exponent is negative. The elegance
of the method, which is equivalent to a decomposition of the
dynamics into eigenmodes, has made it a preferred tool for the
exploration of the properties of synchronization, so that, over
time, it has been extended and applied to a diverse range of
complex networks, underscoring its power and versatility~\cite{Hua09,Sor11,del15,Coo16,del16,Fag20,Gam21,Par22,del22}.

The nature of the MSF has also made it a natural
choice of method to employ when studying systems
whose dynamics is chaotic. In fact, a general classification
scheme has also been presented for the synchronization
behaviour of chaotic systems, based on the positivity
regions of the MSF~\cite{Boc06}. This has shown
that each chaotic system belongs to one of three
classes, which correspond to a vanishing, unbounded
or bounded region of parameters for which their
synchronized state is stable after an initial threshold
in coupling strength. However, notwithstanding
the numerous successes of this method, no systematic
study of the MSF behaviour for periodic systems
had been carried out so far. Note that ensuring
that periodic systems reliably permain in a stable
synchronized state is an important task for numerous
engineering applications where the precision of timing
and the mitigation of jitter, instabilities and
phase noise are of critical relevance. Specific
examples of such situations include synchronizing
AC~power-distribution networks to ensure efficient
and coordinated power delivery~\cite{Mot13,Tan14},
guaranteeing the synchronous operation of digital
communication networks to obtain reliable data transmission~\cite{Kaz06},
and maintaining coherence in clock distribution
trees within electronic devices and circuit boards,
which is essential for optimal performance and
functionality~\cite{Bai22}.

In this article, we close the gap in the synchronizability
of periodic systems by a thorough investigation into their synchronization
dynamics. We use the MSF method introduced by Pecora and Carroll, focusing
on its application to periodic systems, which differs from the predominantly
chaotic systems studied in previous research. Chaotic systems are inherently
challenging to synchronize due to the butterfly effect, which has driven significant
research into their synchronizability. In contrast, it was traditionally believed
that identical periodic systems would synchronize at infinitesimal coupling
strength. However, our findings demonstrate that the MSF behavior for periodic
systems can differ substantially from that of chaotic systems. This difference
stems from the initial value of the MSF, which is equal to the maximum Lyapunov
exponent of the systems. By applying the MSF method to several periodic systems,
we reveal the existence of distinct stable synchronization regions and propose
a classification scheme for periodic systems. We identify two additional classes
of MSF behavior unique to periodic systems. Furthermore, we present examples
where periodic systems do not synchronize at infinitesimal coupling. These
findings can help in expanding the understanding of synchronization in periodic
systems.

Given a connected network of~$N$ diffusively-coupled $d$-dimensional
identical systems with weighted adjecency matrix~$\mathbf W$, its dynamics
is described by the system of equations
\begin{equation}\label{dynamics}
 \dot{\mathbf x}_i = \mathbf F({\mathbf x}_i) - \sigma \sum_{j=1}^N L_{i,j}\mathbf H({\mathbf x}_j)\:,
\end{equation}
where ${\mathbf x}_i$ is a vector with~$d$ components
representing the state of node~$i$, $\mathbf F:\ \mathbb R^d\rightarrow\mathbb R^d$
and $\mathbf H:\ \mathbb R^d\rightarrow\mathbb R^d$ are
vector fields describing the internal dynamics of the
systems and their mutual coupling, respectively, $\sigma$
is the coupling strength and the matrix~$\mathbf L$ is
the graph laplacian of the network, whose elements are
\begin{equation}
 \begin{cases}
            L_{i,i} = \sum_{j=1}^N W_{i,j}\\
            L_{i,j} = -W_{i,j}
           \end{cases}
\end{equation}
Note that, for the sake of brevity, here and in the following
we will omit writing explicit time dependencies, except when we
wish to draw specific attention to them. The definition of the
graph Laplacian in the previous equation makes it a positive
semi-definite zero-row-sum matrix. This means that in
this case, it has one zero eigenvalue ($\lambda_1=0$), while
all the others are positive ($\lambda_i>0$ for $i=2, \dotsc, N$).
Also, its presence in Eq.~\eqref{dynamics} guarantees the existence
of an invariant synchronous solution of the dynamics $\mathbf s(t)$, so that
${\mathbf x}_i(t)=\mathbf s(t)$ for all~$i$. In turn, this allows
one to introduce the \emph{synchronization error vectors}~$\delta{\mathbf x}_i={\mathbf x}_i-\mathbf s$,
which measure the componentwise difference between the state
of each node at a given time and the synchronous solution. If~$\mathbf F$
and~$\mathbf H$ are at least~$C^1$, i.e., if they are continuous
and differentiable, one can linearize them via a vector equivalent
of a first-order Taylor expansion around the synchronous solution,
so that
\begin{equation}
\mathbf F(\mathbf s+ \delta{\mathbf x}_i ) \approx \mathbf F(\mathbf s)+ \hat{J} \mathbf F(\mathbf s) \delta{\mathbf x}_i
\end{equation}
and
\begin{equation}
\mathbf H(\mathbf s+ \delta{\mathbf x}_i ) \approx \mathbf H(\mathbf s)+ \hat{J} \mathbf H(\mathbf s) \delta{\mathbf x}_i\:,
\end{equation}
where $\hat{J}\mathbf F(\mathbf s)$ and $\hat{J}\mathbf H(\mathbf s)$
are the Jacobians of~$\mathbf F$ and~$\mathbf H$, respectively. Then,
substituting ${\mathbf x}_i=\mathbf s+\delta{\mathbf x}_i$ into Eq.~\eqref{dynamics}
leads to
\begin{equation}
 \dot{\mathbf s} + \dot{\delta\mathbf x}_i = \mathbf F(\mathbf s) + \hat{J}\mathbf F(\mathbf s)\delta{\mathbf x}_i-\sigma\sum_{j=1}^NL_{i,j}\left(\mathbf H(\mathbf s) + \hat{J}\mathbf H(\mathbf s)\delta{\mathbf x}_i\right)\:.
\end{equation}
 As $\dot{\mathbf s}=\mathbf F(\mathbf s)$ and $\sum_{j=1}^NL_{i,j}=0$,
 the evolution of the synchronization error vectors simplifies to
\begin{equation}
 \dot{\delta\mathbf x}_i = \hat{J}\mathbf F(\mathbf s)\delta{\mathbf x}_i-\sigma\sum_{j=1}^NL_{i,j}\hat{J} \mathbf H(\mathbf s)\delta{\mathbf x}_i\:.
\end{equation}

Finally, the synchronization error vectors can be decomposed along the directions determined by the eigenvectors
of the Laplacian, which can be conveniently arranged in an orthogonal matrix~$\mathbf V$.
This yields a decomposition of the dynamics into~$N$ decoupled modes~$\boldsymbol\eta_i=\mathbf V^{-1}\delta{\mathbf x}_i$,
whose evolution is given by the set of variational equations
\begin{equation}
 \dot{\boldsymbol\eta}_i = \left(\hat{J}\mathbf F(\mathbf s) - \sigma\lambda_i\hat{J}\mathbf H(\mathbf s)\right)\boldsymbol\eta_i\:.
\end{equation}
Because of the fact that $\lambda_1=0$ and because
of the orthogonality of~$\mathbf V$, the evolution
of these variational equations occurs along the synchronous
solution of the dynamics for $i=1$, and along directions
transverse to it for $i>1$. Then, one can consider
the generic equation
\begin{equation}\label{msfeq}
 \dot{\boldsymbol\eta} = \left(\hat{J}\mathbf F(\mathbf s) - K\hat{J}\mathbf H(\mathbf s)\right)\boldsymbol\eta
\end{equation}
and compute its maximum Lyapunov exponent~$\Lambda$. The dependence of~$\Lambda$
on the generalized coupling strength~$K$ is the Master Stability Function~\cite{Pec98}.
Given a coupling strength~$\sigma$, if the synchronized state is stable for that
value of~$\sigma$, then the MSF is negative for all values of $K=\sigma\lambda_i$
with $i>1$. Note that when $K=0$, the value of the MSF is the maximum Lyapunov exponent
of the uncoupled system. Consequently, for chaotic systems, the MSF has a positive
intercept.

\begin{figure}[t]
 \centering
\includegraphics[width=0.45\textwidth]{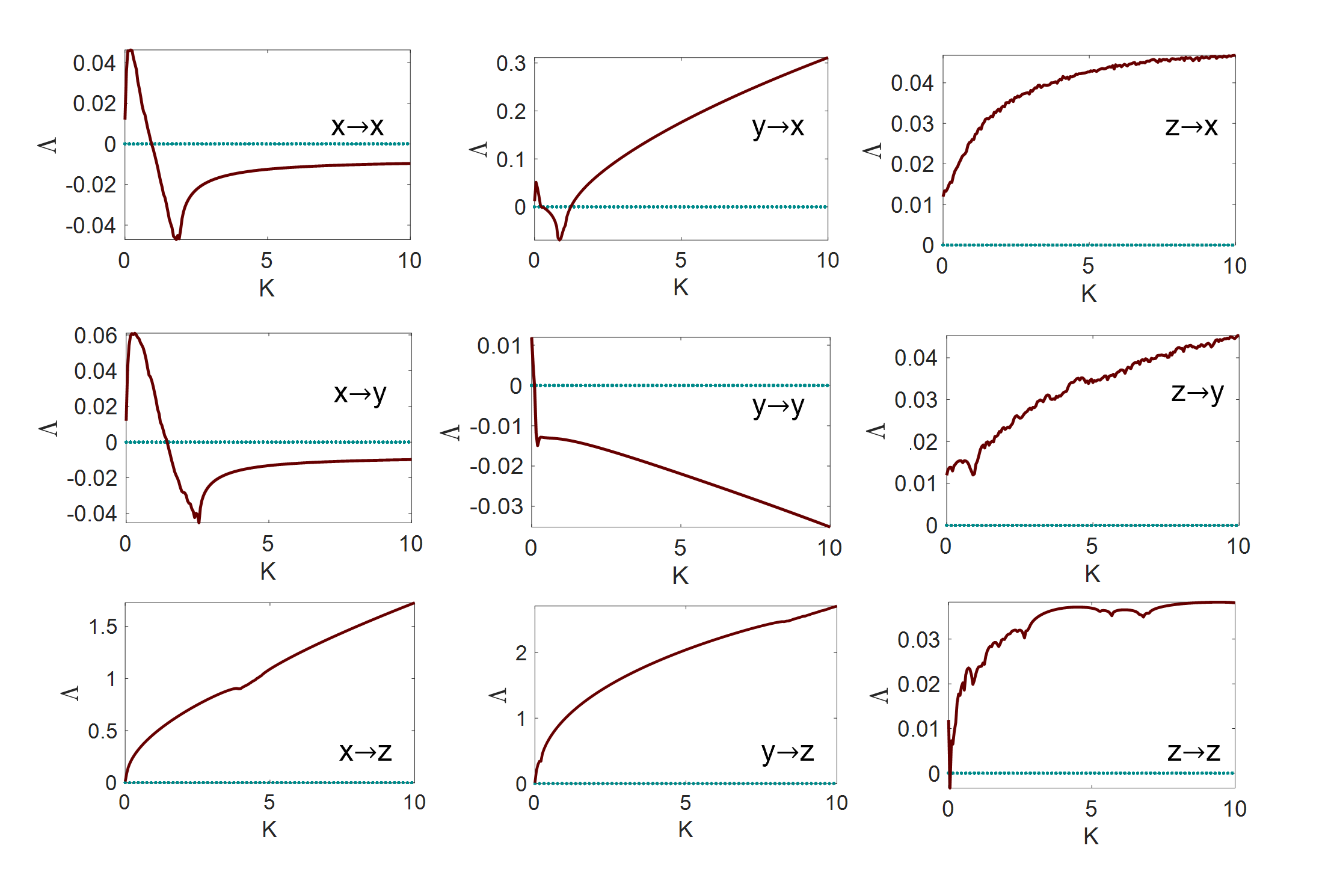}
\caption{\label{FigHRChaos}\textbf{The chaotic
Hindmarsh-Rose system can belong to Classes~I, II and~III.} The
Master Stability Function ($\Lambda$) of the chaotic Hindmarsh-Rose
system, Eq.~\eqref{HREq}, plotted as a function of the generalized
coupling strength $K$, is in Class~III for the couplings $y\rightarrow x$
and $z\rightarrow z$, in Class~II for $x\rightarrow x$, $x\rightarrow y$
and $y\rightarrow y$, and in Class~I for all other choices. Thus,
depending on the choice of the coupling, the Hindmarsh-Rose system
in the chaotic regime can belong to any of the synchronizability
classes. The parameter values are $a=1$, $b=3$, $I=\frac{16}{5}$,
$c=1$, $d=5$, $r=6\times 10^{-3}$, $s=4$ and $x_1=\frac{8}{5}$.}
\end{figure}
Based on the qualitative behaviour of the MSF, a general
classification for the synchronization stability of chaotic
systems was introduced in Ref.~\cite{Boc06}. According to
it, any system belongs to one of the following three classes:
\begin{enumerate}
 \item[] \textbf{Class I} The MSF is positive for all values of~$K$.
 Consequently, synchronization is not stable for any coupling strength.
 \item[] \textbf{Class II} The MSF is negative for an unbounded interval
 of values of~$K$. Consequently, there exists a critical value~$K^\ast$,
 at which the function intersects the horizontal axis and after
 which it is always negative.
 \item[] \textbf{Class III} The MSF is negative in a bounded interval
 of values of~$K$. Consequently, there are two intersection
 points~$K^\ast_1$ and~$K^\ast_2$, such that the MSF is negative
 for $K^\ast_1<K<K^\ast_2$.
\end{enumerate}
As an example of a chaotic system that can belong
to any of the three classes depending on the coupling
between elements, consider the Hindmarsh-Rose model,
whose system of equations describes the spiking and
bursting behaviour of a single neuron~\cite{Hin84}:
\begin{equation}\label{HREq}
 \begin{aligned}
  \dot x &= y - ax^3 + bx^2 - z + I\\
  \dot y &= c - dx^2 - y\\
  \dot z &= -rz + rs(x+x_1)\:.
 \end{aligned}
\end{equation}
To obtain the MSF of the HR system,
first the perturbed equations are derived according
to Eq.~\eqref{msfeq}. Then, the maximum Lyapunov exponent~$\Lambda$
of the perturbed system is calculated as a function
of the generalized coupling strength~$K$, and defined
as the MSF. Choosing $a=1$, $b=3$, $I=\frac{16}{5}$,
$c=1$, $d=5$, $r=6\times 10^{-3}$, $s=4$ and $x_1=\frac{8}{5}$
results in a rich variety of synchronization behaviours
corresponding to the nine possible single-variable couplings
$i\rightarrow j$, with $(i,j)\in\left\lbrace x, y, z\right\rbrace\times\left\lbrace x, y, z\right\rbrace$.
The MSFs of the chaotic HR system are
illustrated in Fig.~\ref{FigHRChaos}. The figure shows
that the MSF can belong to all synchronizability classes
of chaotic systems, namely Class~III for the couplings
$y\rightarrow x$ and $z\rightarrow z$, Class~II for
$x\rightarrow x$, $x\rightarrow y$ and $y\rightarrow y$,
and Class~I for all other couplings.

The situation is subtly different when one considers
periodic systems. In fact, if an isolated system supports
a periodic orbit, its corresponding maximum Lyapunov
exponent is~0. This means that the MSF in the case of
periodic systems does not start from a strictly positive
value, but rather it starts from~0. This has two immediate
consequences. First, there is always at least one point
where the MSF vanishes, namely $K=0$. Second, an initial
discrimination for the synchronizability of a coupled
network is determined by the sign of the (right-hand)
derivative of the MSF at~0. This suggests the possibility
that the properties of synchronizability of a network
of periodic oscillators are actually more complex than
those of a network of chaotic ones.

\begin{figure}[t]
 \centering
\includegraphics[width=0.45\textwidth]{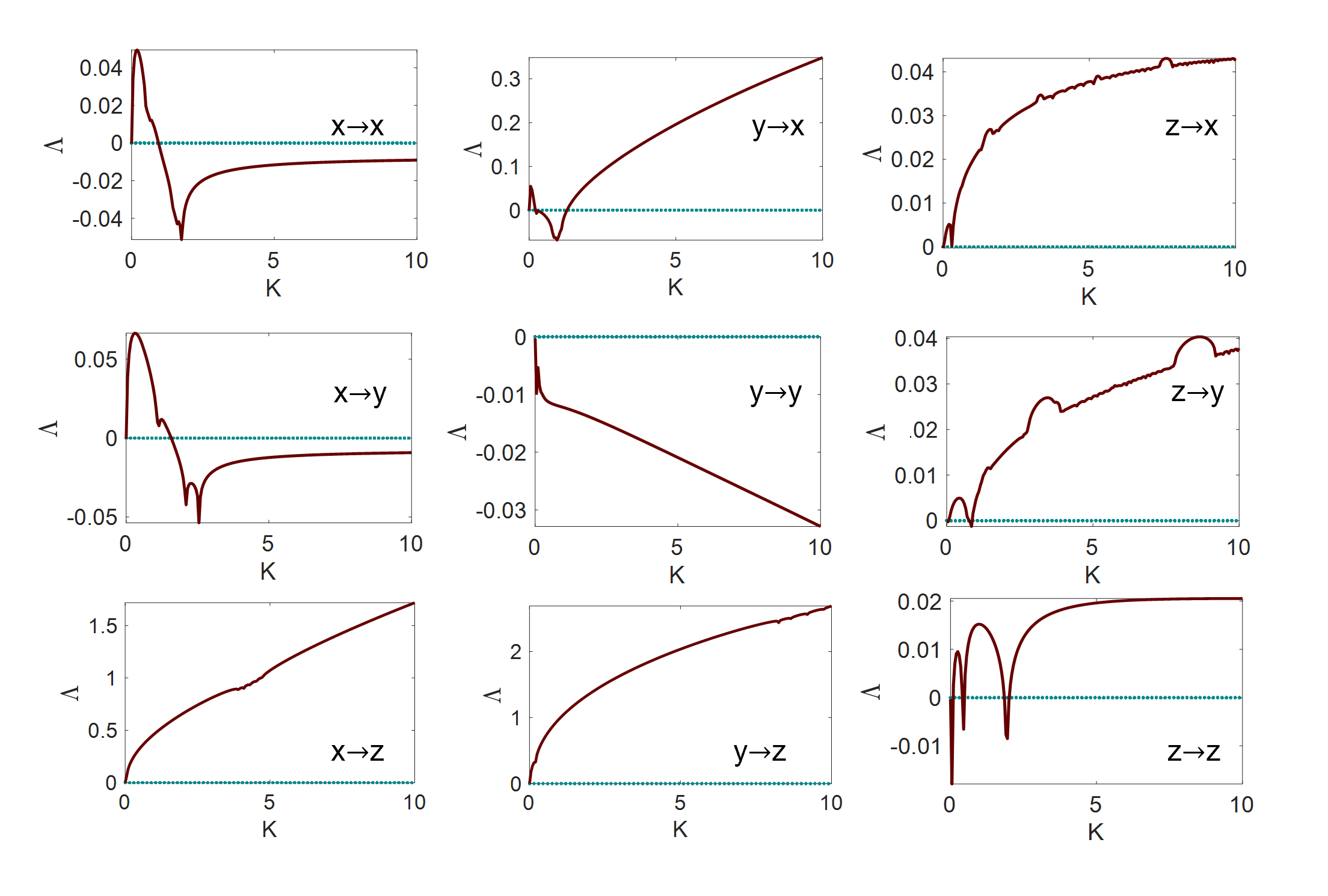}
\caption{\label{FigHRPer}\textbf{New classes
of synchronizability for the periodic Hindmarsh-Rose system.}
The Master Stability Function ($\Lambda$) of the periodic Hindmarsh-Rose
system, Eq.~\eqref{HREq}, plotted as a function of the generalized
coupling strength~$K$, is always positive for the couplings $x\rightarrow z$,
$y\rightarrow z$ and $z\rightarrow x$, it has bounded intervals
of negative values for $y\rightarrow x$ and $z\rightarrow y$,
it is negative in an unbounded range for $x\rightarrow x$, $x\rightarrow y$
and $y\rightarrow y$, and it has multiple regions of negativity
for $z\rightarrow z$. Note that the $z\rightarrow z$ coupling
results in the system belonging to a synchronization class that
is exclusive of periodic systems. Also note that, in contrast
with conventional belief, a minimum coupling strength is required
to achieve synchronization in many cases. The parameter values
are $a=1$, $b=3$, $I=\frac{16}{5}$, $c=1$, $d=5$, $r=5.6\times 10^{-3}$,
$s=4$ and $x_1=\frac{8}{5}$.}
\end{figure}
To confirm the correctness of this consideration, we computed
the MSF for the Hindmarsh-Rose model, using the same parameter
values as before, except for~$r$, which we imposed to be equal
to~$5.6\times 10^{-3}$. This choice ensures that dynamics of
the individual systems is periodic. It should
be noted that the steps one has to follow in order to compute
the MSF for a periodic system are the same as would be taken
in the case of a chaotic system. The only difference is that
the periodic synchronous solution is used to obtain the perturbed
linear equation. The MSF results, shown in Fig.~\ref{FigHRPer},
demonstrate an even broader range of behaviours than observed
in the chaotic version of the model. In fact, the couplings $x\rightarrow z$,
$y\rightarrow z$ and $z\rightarrow x$ result in a MSF that is
always positive, the couplings $y\rightarrow x$, $z\rightarrow y$
and $z\rightarrow z$ yield well-defined ranges of negativity,
and the couplings $x\rightarrow x$, $x\rightarrow y$ and $y\rightarrow y$
result in unbounded regions of negative values for the MSF. Therefore,
the fundamental effect of the periodicity of the system is on
the MSF of the couplings $z\rightarrow y$ and $z\rightarrow z$,
with the former that now features a bounded negative region and
the latter that includes several negative regions. Moreover,
even though $x\rightarrow x$, $x\rightarrow y$ and $y\rightarrow y$
all correspond to unbounded regions of negativity, only in the
case of $y\rightarrow y$ does the region start at $K=0$. Effectively,
one could say that the fact that the MSF vanishes at 0 has split
Class~II into two new classes: if the derivative at~0 is negative,
then the unbounded region of stability starts at~0; if, instead,
it is positive, then the interval starts at a value $K^\ast>0$.
Similarly, $y\rightarrow x$, $z\rightarrow y$ and $z\rightarrow z$
produce a finite region of negative values, which, however, only
starts at~0 for the $z\rightarrow z$ coupling. Thus, also Class~III
undergoes a split that depends on the sign of the derivative
at~0, akin to that of Class~II. Note that the $z\rightarrow z$
coupling actually produces multiple separate intervals for which
the MSF is negative. However, when classifying the stability
of synchronized states, one is generally only interested in the
interval after the first threshold for stability, which, in this
case, is~0.

\begin{figure}[t]
 \centering
\includegraphics[width=0.45\textwidth]{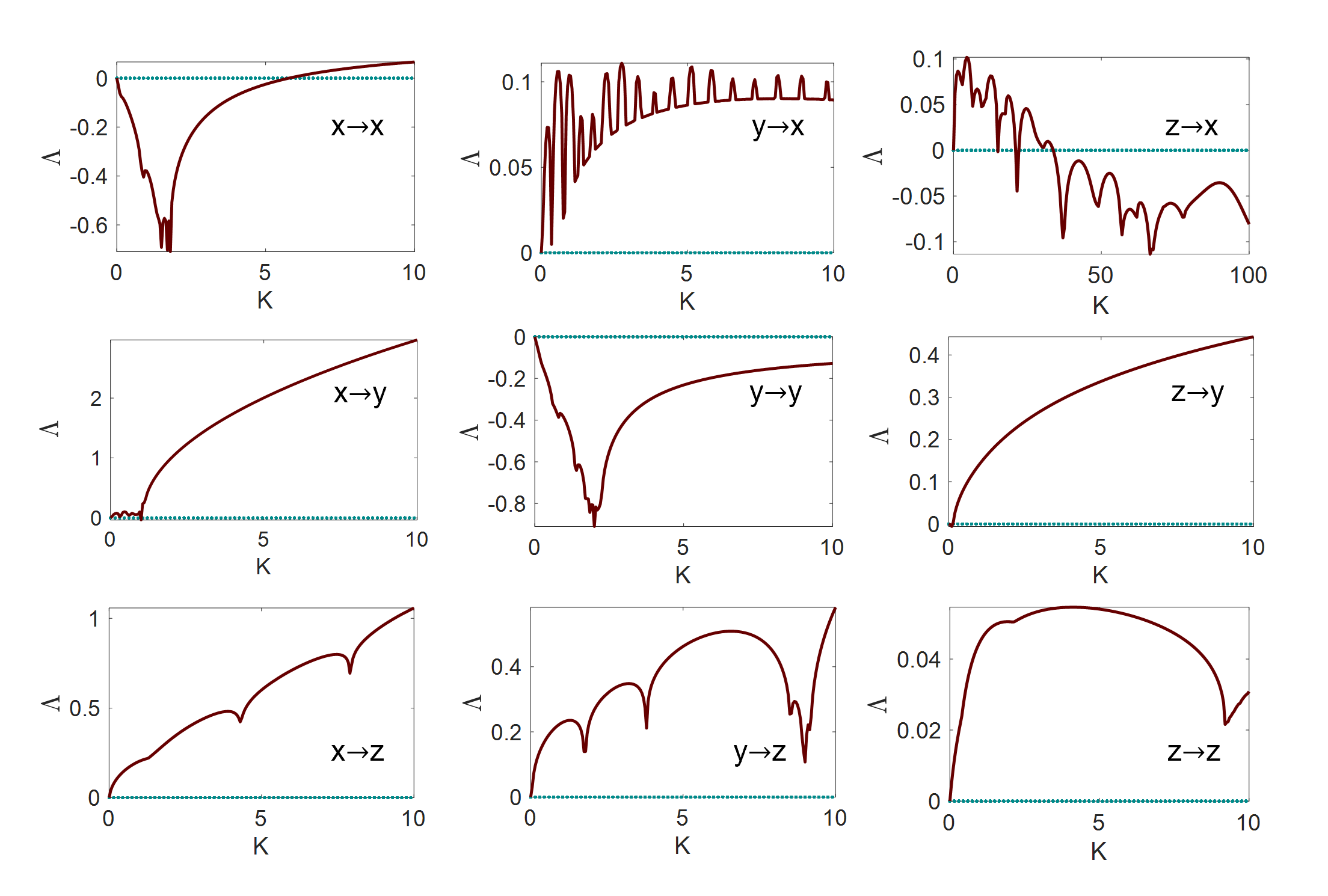}
\caption{\label{FigRosPer}\textbf{New classes
of synchronizability for the periodic Rössler oscillator.} The
Master Stability Function ($\Lambda$) of the periodic Rössler
oscillator, Eq.~\eqref{RosEq}, plotted as a function of the generalized
coupling strength~$K$, is always positive for all couplings except
$x\rightarrow x$, which results in a bounded negative region,
$y\rightarrow y$, which yields an unbounded negative region, and
$z\rightarrow x$, which causes the appearance of multiple negative
intervals. Note how the $z\rightarrow x$ coupling induces a minimum
coupling strength after which synchronization is always stable,
whereas with the $x\rightarrow x$ coupling a maximum coupling
strength emerges after which synchronization is never stable,
contrary to the received wisdom about periodic systems. The parameter
values are $a=0.161$, $b=0.2$ and $c=9$.}
\end{figure}
To further explore this phenomenology,
we studied a network of Rössler oscillators~\cite{Ros76},
whose evolution is given by the sytem
\begin{equation}\label{RosEq}
 \begin{aligned}
  \dot x &= -y -z\\
  \dot y &= x + ay\\
  \dot z &= b + (x-c)z\:.
 \end{aligned}
\end{equation}
To ensure periodic dynamics, we chose the parameter
values $a=0.161$, $b=0.2$ and $c=9$. The calculation
of the MSF for all possible single-variable couplings,
illustrated in Fig.~\ref{FigRosPer}, shows that in
all cases except $x\rightarrow x$, $y\rightarrow y$
and $z\rightarrow x$ the maximum Lyapunov exponent
remains positive for all $K>0$. The $y\rightarrow y$
coupling results in an unbounded negative region starting
at~0. For $x\rightarrow x$, the MSF is negative only
in a range $0<K<k^\ast$. Finally, the $z\rightarrow x$
causes a situation similar to the Hindmarsh-Rose model
with $z\rightarrow z$ coupling, with the appearance
of multiple finite intervals of stable synchronization,
the first of which starts at a positive~$K^\ast_1$.
This confirms the occurrence of a split in Class~III,
dependent on the sign of the derivative of the MSF
at~0: for negative derivatives one obtains a stable
region $0<K<k^\ast$, whereas for positive derivatives
stability happens for $K^\ast_1<K<K^\ast_2$, with $K^\ast_1>0$.

A similar range of classes of synchronizability
also characterizes the behaviour of the Lorenz
system, which can be always unstable, always stable
with a vanishing or non-zero threshold, or with
a bounded region of stability starting at a positive
coupling strength (see Supplementary Material).

\begin{figure}[t]
 \centering
\includegraphics[width=0.45\textwidth]{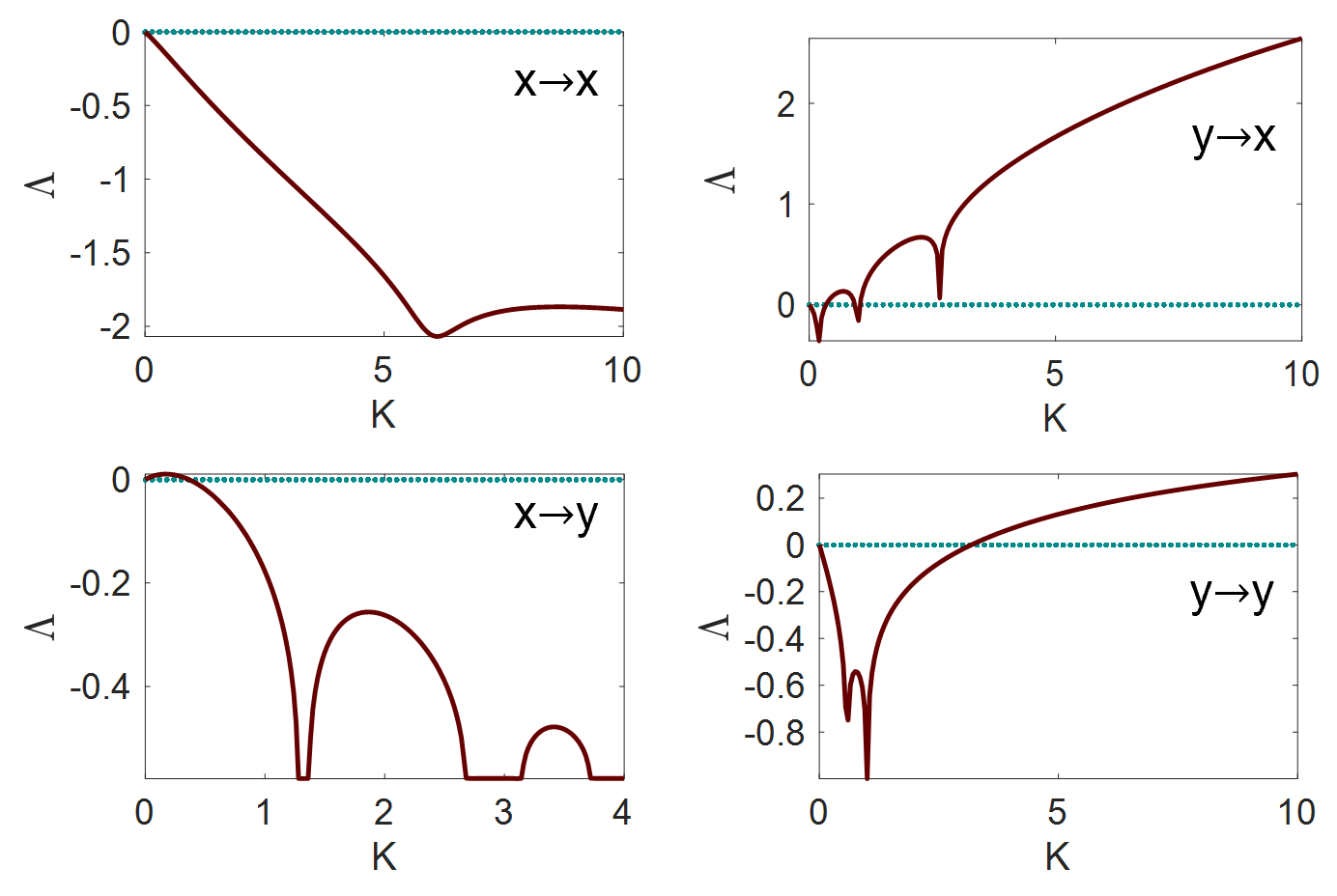}
\caption{\label{FigBruPer}\textbf{New classes
of synchronizability for the periodic Brusselator.} The Master
Stability Function ($\Lambda$) of the periodic Brusselator, Eq.~\eqref{BruEq},
plotted as a function of the generalized coupling strength~$K$,
is always negative for $x\rightarrow x$ coupling, negative after
a threshold for $x\rightarrow y$, negative before a threshold
for $y\rightarrow y$, and has multiple regions of negativity for
$y\rightarrow x$. Note that the synchronization class corresponding
to the $y\rightarrow x$ coupling is exclusive of periodic system.
Also, the $x\rightarrow y$ coupling requires a minimum coupling
strength to achieve stable synchronization, whereas the $y\rightarrow x$
and $y\rightarrow y$ couplings induce a maximum coupling strength,
after which synchronization ceases to be stable, in contrast with
the current assumptions about periodic systems. The parameter values
are $a=1$ and $b=3$.}
\end{figure}
To check whether the dimensionality of the oscillators
plays a role in the emergence of the new synchronizability
classes, we studied several 2-dimensional systems. We
found the most diverse behaviour is exhibited by the Brusselator
system~\cite{Pri67}, which is a mathematical model for
autocatalytic chemical reactions described by the following
system:
\begin{equation}\label{BruEq}
 \begin{aligned}
  \dot x &= a + x^2y - \left( b-1\right)x\\
  \dot y &= bx - x^2y\:.
 \end{aligned}
\end{equation}
Choosing $a=1$ and $b=3$, we obtain the MSF represented
in Fig.~\ref{FigBruPer}. Since the model is 2-dimensional,
we have only 4~possible single-variable couplings. Notably,
each of them produces a different synchronizability behaviour:
$x\rightarrow x$ results in the MSF being always negative
for $K>0$, $x\rightarrow y$ causes the appearance of an
unbounded region of stability after a $K^\ast>0$, and $y\rightarrow x$
and $y\rightarrow y$ corresponds to a bounded stability
region for $0<K<K^\ast$, which, in the case of $y\rightarrow x$,
is followed by a second one.

\begin{figure}[t]
 \centering
\includegraphics[width=0.45\textwidth]{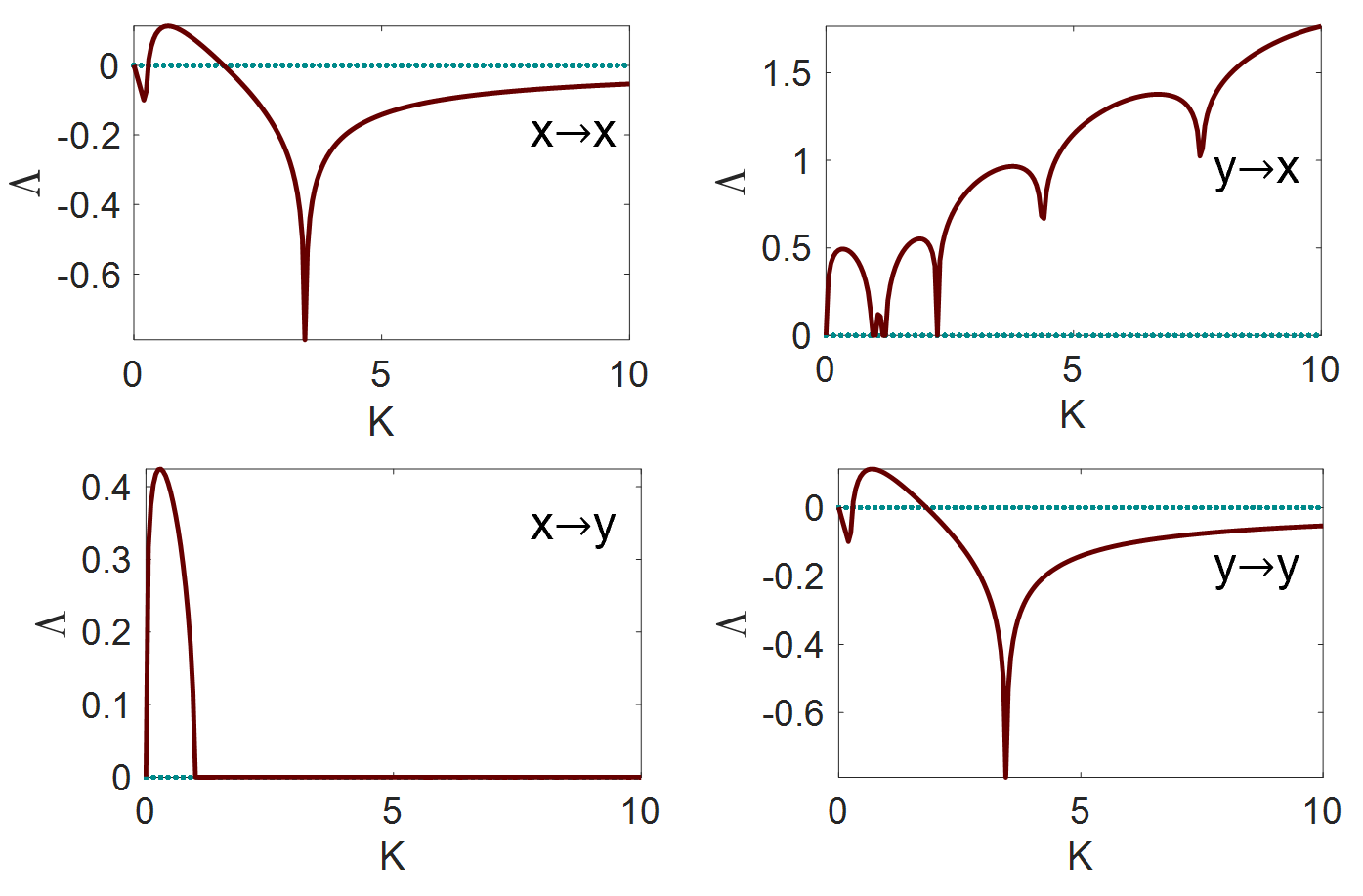}
\caption{\label{FigDufPer}\textbf{New classes
of synchronizability for the unforced undamped Duffing oscillator.}
The Master Stability Function ($\Lambda$) of the unforced undamped
Duffing oscillator, defined in Eq.~\eqref{DufEq}, plotted as a
function of the generalized coupling strength~$K$, has multiple
negative regions for self-couplings, and it is always positive
otherwise. Thus, the self-couplings correspond to a new synchronizability
class, which is exclusive for periodic systems, and the others
contradict the current general assumption that periodic systems
synchronize in a stable way for any positive coupling strength.}
\end{figure}

A slightly less rich behaviour is offered by the unforced
undamped Duffing oscillator~\cite{Duf921}, whose dynamics
is given by the system
\begin{equation}\label{DufEq}
 \begin{aligned}
  \dot x &= y\\
  \dot y &= x - x^3\:.
 \end{aligned}
\end{equation}
In fact, its MSF is either always positive
for all $K>0$, when the coupling is $x\rightarrow y$
or $y\rightarrow x$, or it has multiple intervals
of negativity, with the first one starting
at~0, when the coupling is $x\rightarrow x$
or $y\rightarrow y$ (Fig.~\ref{FigDufPer}).

Similar behaviours are observed in several other
2-dimensional periodic systems that we have
systematically studied, namely the Lotka-Volterra
model, the FitzHugh–Nagumo model, the van~der~Pol
oscillator, the cabbage system and the Stuart–Landau
oscillator (see Supplementary Material). In all
these cases, we have found the appearance of different
synchronizability classes, including split ones.

\begin{figure}[t]
 \centering
\includegraphics[width=0.45\textwidth]{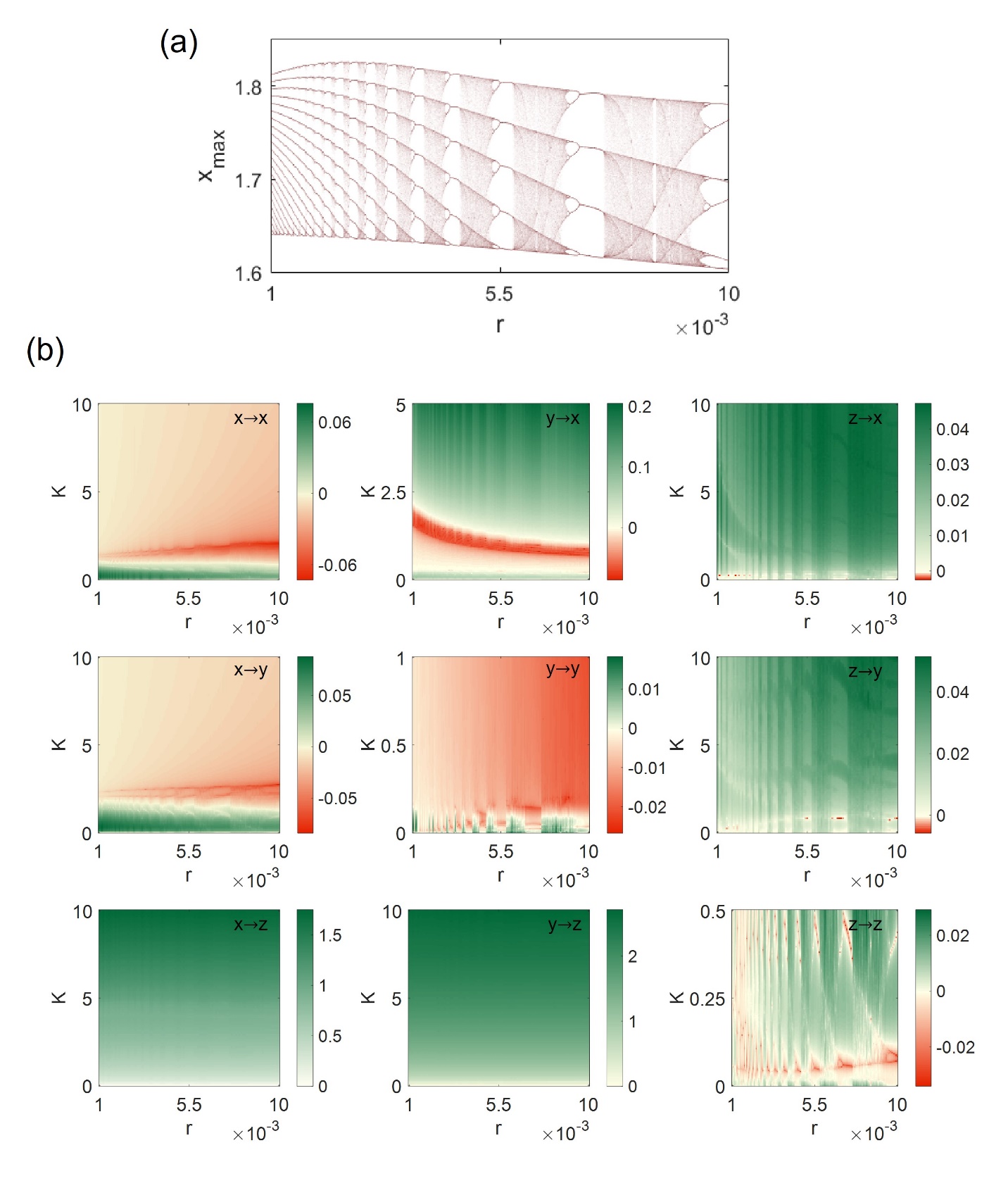}
\caption{\label{FigHRSweep}\textbf{System
parameters can cause a switch in synchronizability properties.}
(a) The bifurcation diagram of the Hindmarsh-Rose system, Eq.~\eqref{HREq},
with $a=1$, $b=3$, $I=\frac{16}{5}$, $c=1$, $d=5$, $s=4$ and
$x_1=\frac{8}{5}$, plotting the largest value of~$x$ ($x_{max}$)
for different values of~$r$, shows numerous transitions between
periodic and chaotic behaviour. (b) The MSF ($\Lambda$), plotted
for different values of~$r$ and $K$ in different coupling
schemes, shows that in some cases, such as $y\rightarrow y$,
the stable synchronizability region can change its profile.}
\end{figure}
Note that, in all the cases considered,
it is to be expected that, for a given
type of coupling, different parameter values
will result in a different synchronizability
profile. As an example, consider again
the Hindmarsh-Rose system. Its bifurcation
diagram, illustrated in Fig.~\ref{FigHRSweep}
for the same parameter values as used before
and using~$r$ as control, shows the existence
of multiple transitions between periodic
and chaotic dynamics. Studying the MSF
for different values of~$r$, one can observe
that its qualitative behaviour remains
unchanged under $x\rightarrow x$, $y\rightarrow x$,
$x\rightarrow y$, $x\rightarrow z$ and
$y\rightarrow z$ couplings. However, for
other coupling schemes, the behaviour
of the MSF depends on the value of~$r$.
Most clearly, in the $y\rightarrow y$
coupling, the MSF can be either always
negative, or it can first take on positive
values and then turn permanently negative,
as synchronization becomes stable~(Fig.~\ref{FigHRSweep}),
explicitly showing how a parameter change
can switch the sign of the derivative of
the MSF at~0 and, consequently, alter
the stability properties of the synchronized
state.

In summary, we have explicitly shown how the Master
Stability Function for periodic networked systems can
have a wealth of different behaviours. In particular,
Class~II and Class~III for chaotic systems, corresponding
to unbounded and bounded regions of negativity of the
MSF, respectively, split each into two different classes
when the systems considered are periodic. This symmetry breaking
is caused by the fact that, when the coupling is~0,
the MSF is the largest Lyapunov exponent of the uncoupled
system, which, in the periodic case, is~0. Thus, there
is always at least one point at which the MSF touches
the horizontal axis, namely the point at~0. In turn,
this means that it is always possible that $K=0$ is
a threshold value for the MSF, whether the unique one,
like in Class~II, or the lower one, like in Class~III.
The sign of the derivative of the MSF at the origin
determines however whether, for very small values of
the coupling, the function is negative or positive.
Therefore, if the derivative is negative, a region of coupling
strengths for which the synchronous state is stable
starts immediately, whereas synhronization is otherwise
unstable for low coupling strengths. Based on
these considerations, and in analogy with chaotic systems,
we propose the following classification of synchronizability
of periodic systems:
\begin{enumerate}
 \item[] \textbf{Class I} The MSF is positive for all $K>0$.
 Thus, synchronization is never stable for any coupling strength.
 \item[] \textbf{Class II} The MSF is negative for all $K>0$.
 This class, corresponding to Class~II of chaotic systems with
 $K^\ast=0$, contains systems whose synchronous state is stable
 for any coupling strength.
 \item[] \textbf{Class III} The MSF is negative for $0<K<K^\ast$.
 This class, corresponding to Class~III of chaotic systems with $K^\ast_1=0$
 and $K^\ast_2=K^\ast$, contains systems whose synchronous state
 is stable only for non-zero couplings smaller than a threshold.
 \item[] \textbf{Class IV} The MSF is negative for $K>K^\ast$,
 with $K^\ast>0$. This class is exclusive to periodic systems. In
 fact, even though it resembles Class~II of chaotic systems, it is
 to be noted that, in that case, the first point at which the MSF
 vanishes must have negative derivative, whereas here the derivative
 at the first root of the MSF is positive.
 \item[] \textbf{Class V} The MSF is negative in a range $K^\ast_1<K<K^\ast_2$.
 Similar to the previous case, this class is typical of periodic systems even though
 it resembles class~III of chaotic ones.
\end{enumerate}
Note that, since the value of the MSF at~0 is always~0
for periodic systems, this classification is exhaustive,
because of the dependence of the new classes on the positivity
of the derivative at~0.

Additionally, our results challenge some of the received wisdom about periodic
systems. In fact, it was generally believed that periodic systems can
achieve a stable synchronized state even for small coupling strengths.
However, we have demonstrated that in some cases, such as those
falling into Class~III and Class~V, too strong a coupling can destroy
the stability of synchronization. Even more to the point, the existence
of Class~IV and, again, Class~V shows that, sometimes, there is
indeed even a non-zero \emph{lower} threshold for stability. Moreover,
Classes~III and~V have some fascinating implications, especially
when they feature multiple stability regions with an an unbounded final
one. In fact, in these cases, systems have to admit a synchronous
state that is definitively stable for large enough coupling. However,
at the same time, the stability of synchronization may be temporarily
lost as the coupling strength increases, before reaching a final
threshold. While this is not too much surprising in chaotic systems,
these behaviours, which we have clearly identified, were believed
not to occur in periodic systems, highlighting the value ot the Master
Stability Function approach in studying the synchronization of nonlinear
systems, regardless of their nature.

This work is partially funded by the Centre for Nonlinear
Systems, Chennai Institute of Technology, India, vide funding
number CIT/CNS/2024/RP/012. L.M. acknowledges
support of the ``Hundred Talents'' program of the University
of Electronic Science and Technology of China, of the ``Outstanding
Young Talents Program (Overseas)'' program of the National
Natural Science Foundation of China, and of the talent programs
of the Sichuan province and Chengdu municipality. C.I.D.G. acknowledges
funding from the Bulgarian Ministry of Education and Science,
under project number BG-RRP-2.004-0006-C02. S.B.
acknowledges support from the project n.PGR01177
of the Italian Ministry of Foreign Affairs and International Cooperation.

\end{document}


\title{Periodic systems have new classes of synchronization stability\\Supplementary Material}
\date{\today}

\maketitle

\section{Introduction}
We present here the Master Stability Functions of several other
2-dimensional and 3-dimensional periodic systems, illustrating
how they can fall into any of the 5~classes of synchronizability discussed in the conclusions of the main text.

\section{The Lotka-Volterra model}
\begin{figure}[p]
 \centering
\includegraphics[width=0.6\textwidth]{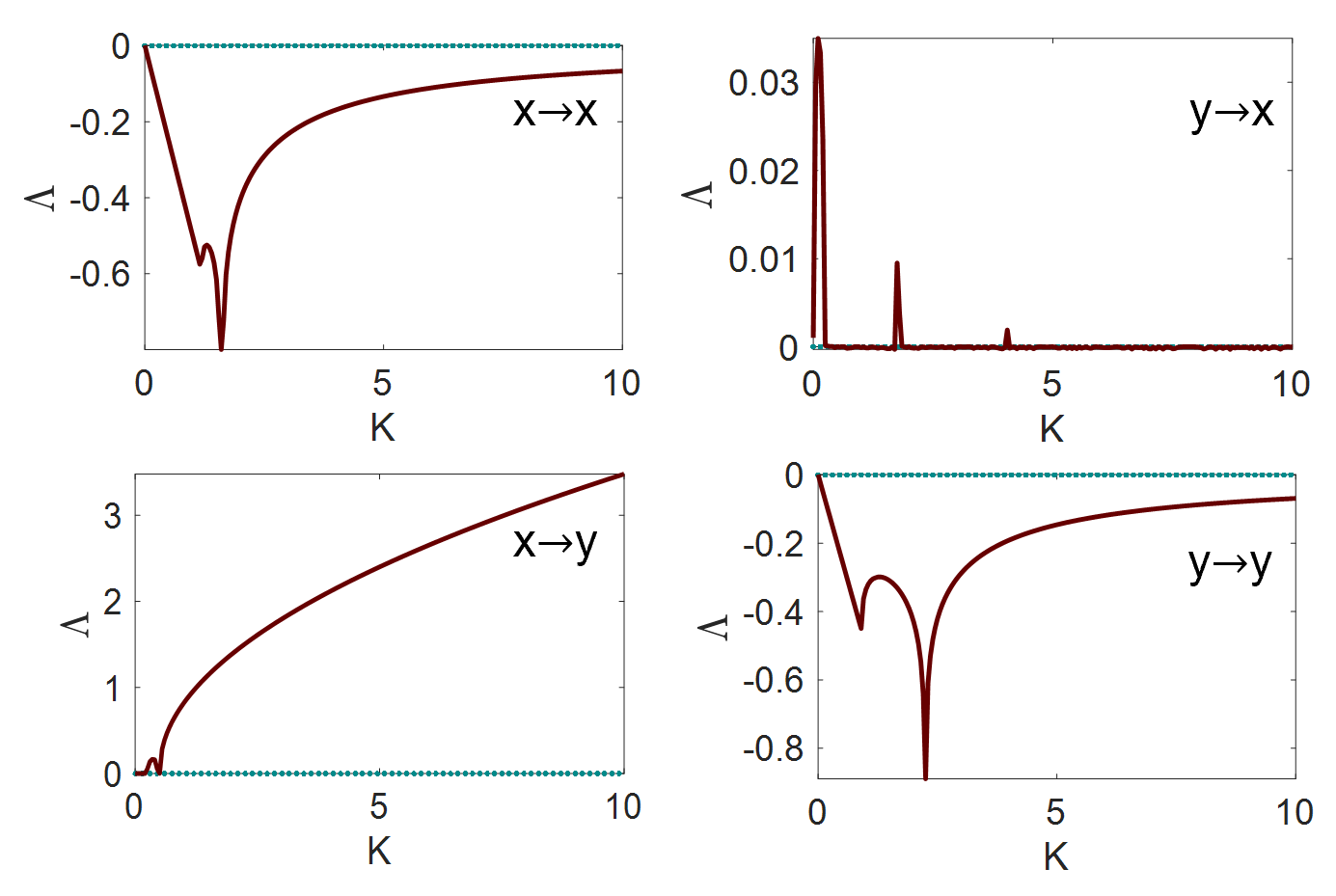}
\caption{\label{FigLVPer}\textbf{Synchronization
in the periodic Lotka-Volterra model can be always unstable.} The
Master Stability Function ($\Lambda$) of the Lotka-Volterra model,
defined in Eq.~\eqref{LVEq}, as a function of the generalized coupling
strength~$K$ is always positive for the $x\rightarrow y$ and $y\rightarrow x$
couplings, showing that synchronization is always unstable in these
cases. The parameter values are $a=\frac{2}{3}$, $b=\frac{4}{3}$,
$c=1$ and $d=1$.}
\end{figure}

The Lotka-Volterra system is a prototypical
predator-prey ecological system, describing
the interaction between two species~\cite{Lot920},
described by the system
\begin{equation}\label{LVEq}
 \begin{aligned}
  \dot x &= ax - bxy\\
  \dot y &= cxy - dy\:.
 \end{aligned}
\end{equation}
Choosing $a=\frac{2}{3}$, $b=\frac{4}{3}$,
$c=1$ and $d=1$ results in a synchronous state
that is always stable (Class~II) for self-couplings
and always unstable (Class~I) otherwise (Fig.~\ref{FigLVPer}).

\section{The FitzHugh–Nagumo model}
\begin{figure}[p]
 \centering
\includegraphics[width=0.6\textwidth]{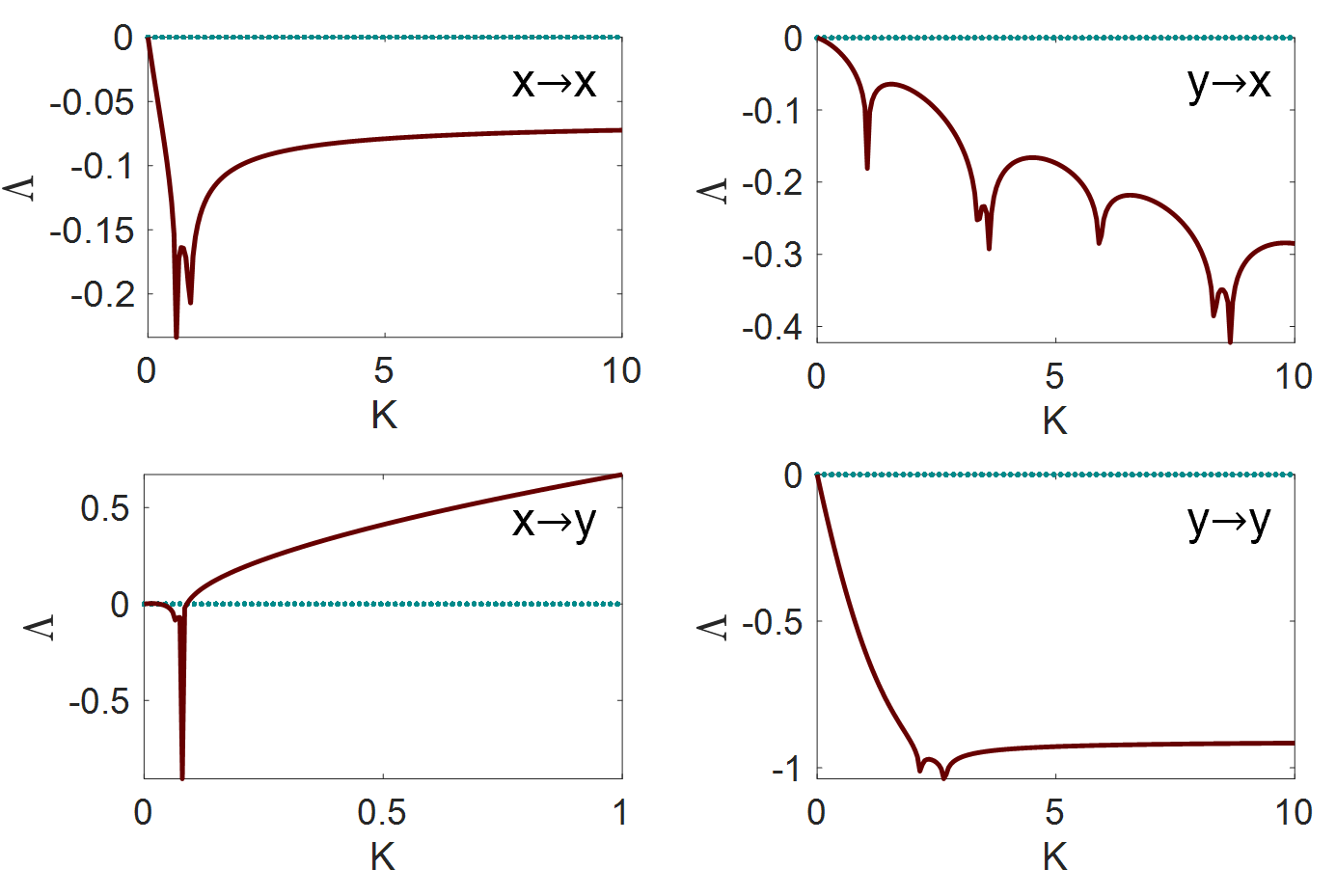}
\caption{\label{FigFHNPer}\textbf{Synchronizability
classes of the periodic FitzHugh-Nagumo model.} The Master Stability
Function ($\Lambda$) of the FitzHugh–Nagumo model, defined in Eq.~\eqref{FHNEq},
plotted as a function of the generalized coupling strength~$K$, is negative
within a bounded range of coupling strengths starting at~0 for the $x\rightarrow y$
coupling, and always negative otherwise. The parameter values are $I=0.5$,
$c=12.5$, $a=0.7$ and $b=0.8$.}
\end{figure}

The FitzHugh–Nagumo model is a simplified
two-dimensional system that can model the
action potential and spiking behaviour of
the neuron cell as a relaxation oscillator~\cite{Fit61,Nag62}.
The equations defining this system are
\begin{equation}\label{FHNEq}
 \begin{aligned}
  \dot x &= x - \frac{x^3}{3} - y + I\\
  c \dot y &= x + a - by\:.
 \end{aligned}
\end{equation}
Imposing the parameter choice $I=0.5$,
$c=12.5$, $a=0.7$ and $b=0.8$, the system
is in Class~II, with its MSF always negative
for any positive coupling strength for
all variable couplings except $x\rightarrow y$,
for which it is in Class~III (Fig.~\ref{FigFHNPer}).

\section{The van~der~Pol oscillator}
\begin{figure}[p]
 \centering
\includegraphics[width=0.6\textwidth]{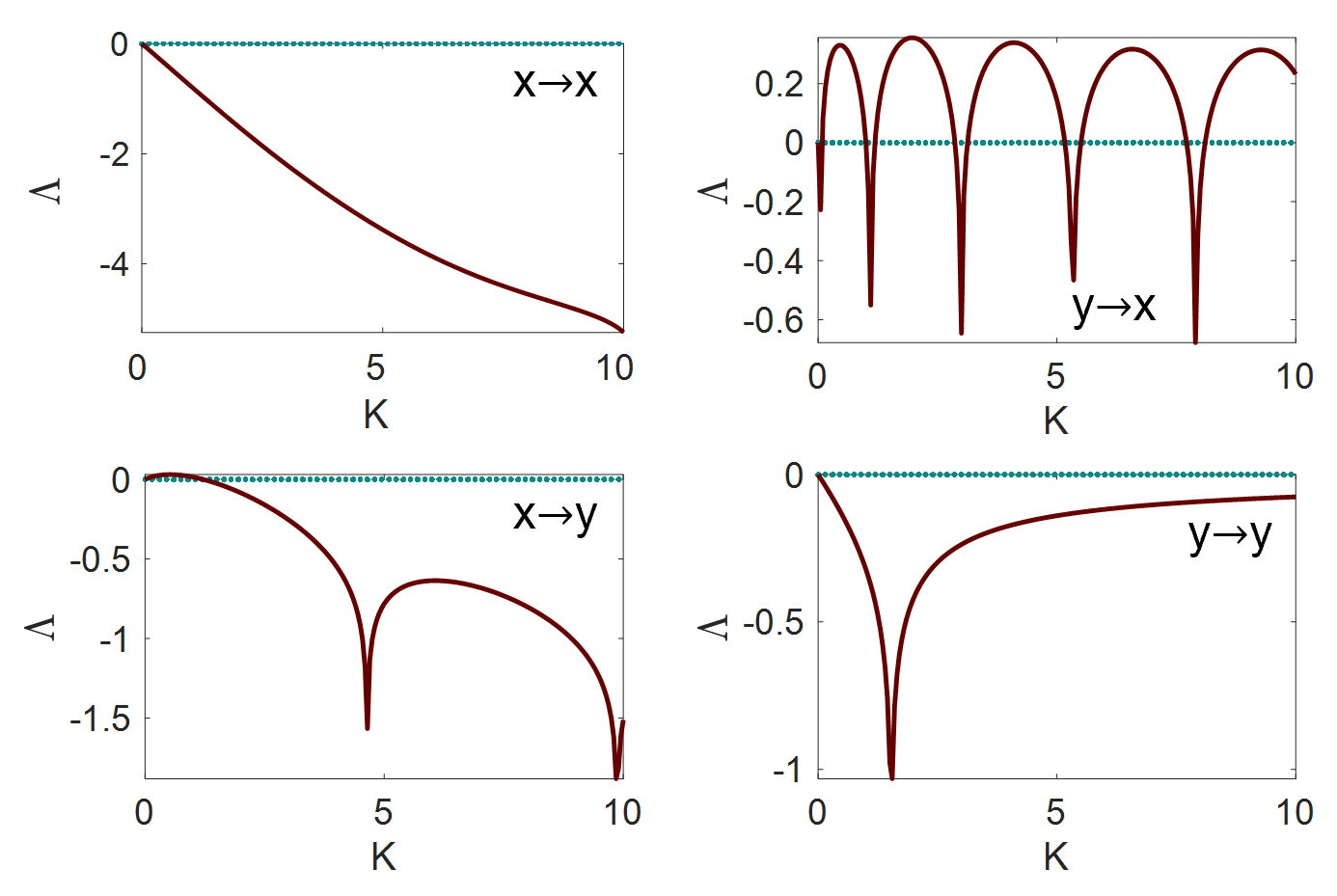}
\caption{\label{FigvdPPer}\textbf{New classes
of synchronizability for the periodic van~der~Pol oscillator.}
The Master Stability Function ($\Lambda$) of the van~der~Pol oscillator,
defined in Eq.~\eqref{vdPEq}, plotted as a function of the generalized
coupling strength~$K$, is always negative for self-couplings,
it is negative after a threshold for $x\rightarrow y$, and it
has multiple ranges of coupling strength for which it is negative
for $y\rightarrow x$. Note how the results for $x\rightarrow y$
and $y\rightarrow x$ couplings challenge the zero-synchronizability
expectation for periodic systems.}
\end{figure}

The van~der~Pol oscillator, is a 2-dimensional
non-conservative oscillating system with a nonlinear
damping term, which, in fact, inspired the development
of the FitzHugh-Nagumo model~\cite{van920}. The
equations that describe its dynamics are
\begin{equation}\label{vdPEq}
 \begin{aligned}
  \dot x &= y\\
  \dot y &= a\left( 1-x^2\right)y - x\:.
 \end{aligned}
\end{equation}
With the choice $a=3.5$, we obtain a system
that can be in one of three classes, depending
on the coupling. Specifically, for self-couplings
it is in Class~II, with an always-stable synchronous
state, for $x\rightarrow y$ coupling it is
in Class~IV, with stability only after a threshold,
and for $y\rightarrow x$ coupling there is the appearance of multiple intervals
of stability (Fig.~\ref{FigvdPPer}).
\begin{figure}[p]
 \centering
\includegraphics[width=0.6\textwidth]{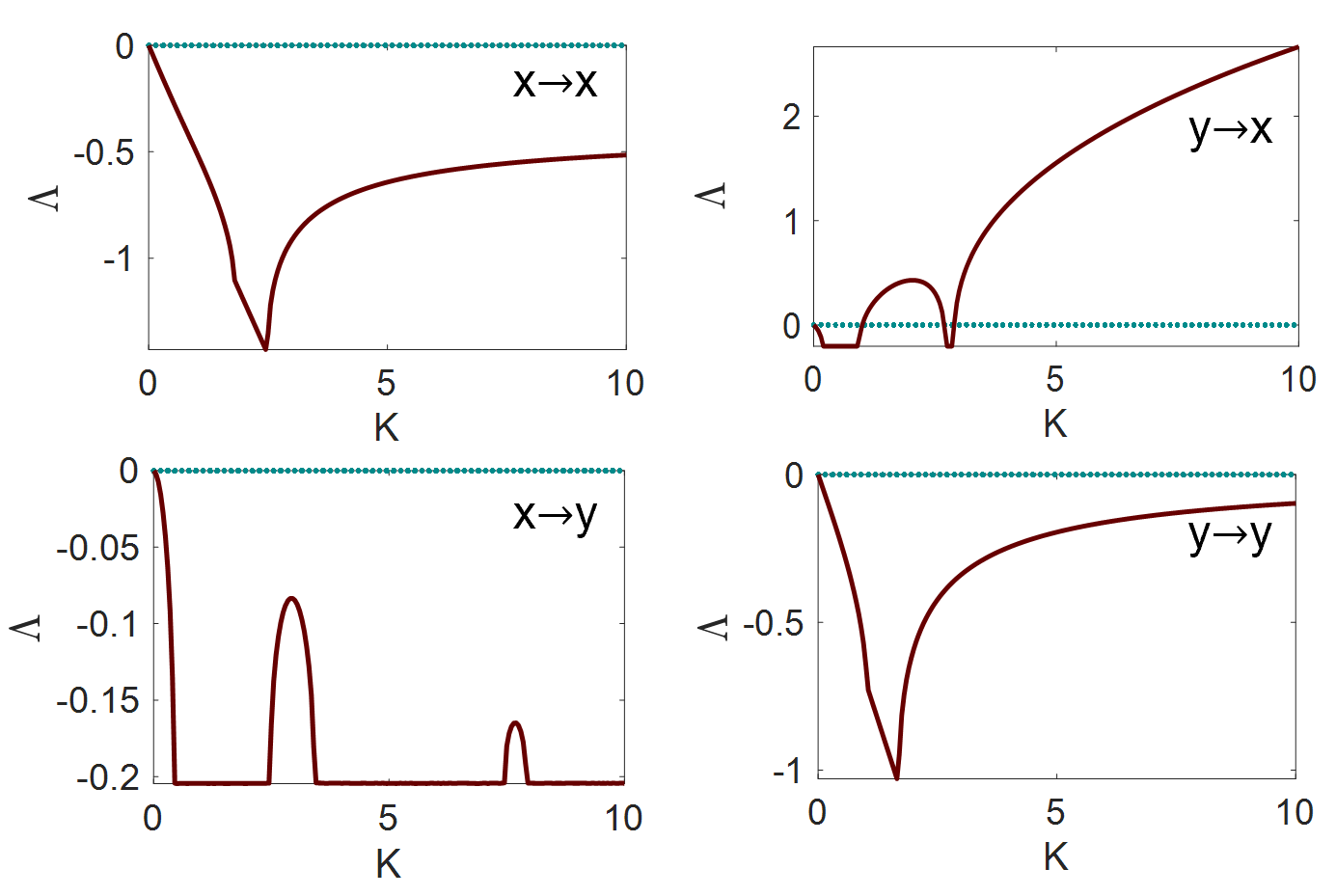}
\caption{\label{FigCabPer}\textbf{New classes
of synchronizability for the periodic cabbage system.} The Master
Stability Function ($\Lambda$) of the cabbage system, defined
in Eq.~\eqref{cabEq}, plotted as a function of the generalized
coupling strength~$K$, is always negative for all couplings except
$y\rightarrow x$, where it has two bounded intervals, one starting
at~0, in which it is negative. This contrasts with the belief
that any coupling, no matter how small will induce stable synchronization
in periodic systems.}
\end{figure}
\begin{figure}[p]
 \centering
\includegraphics[width=0.6\textwidth]{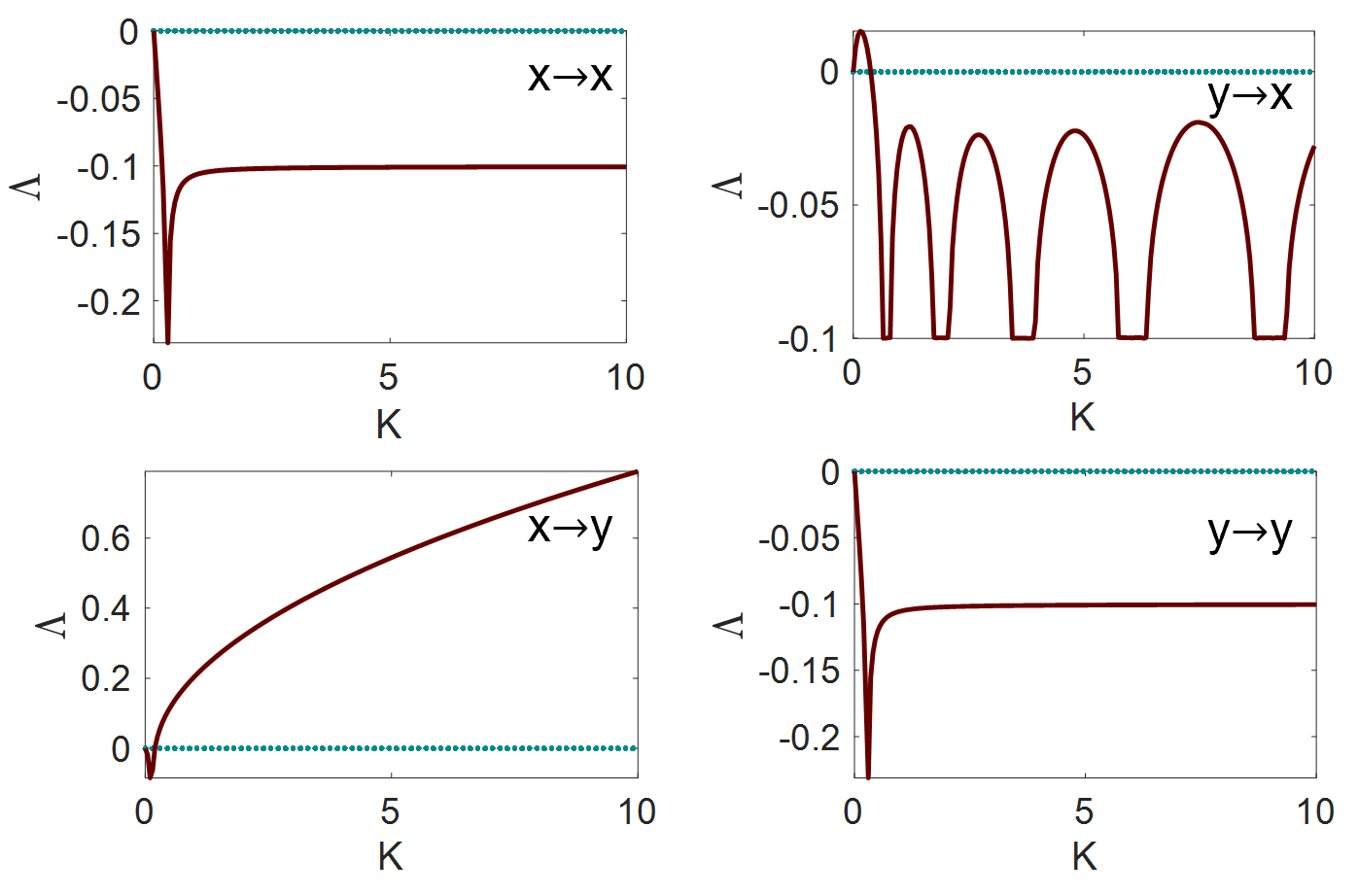}
\caption{\label{FigSLPer}\textbf{New classes
of synchronizability for the Stuart-Landau oscillator.} The Master
Stability Function ($\Lambda$) of the Stuart-Landau oscillator,
defined in Eq.~\eqref{SLEq}, plotted as a function of the generalized
coupling strength~$K$, has three synchronizability classes. The
MSF is always negative for all self-couplings, it is negative
before a threshold for $x\rightarrow y$, and it is negative after
a threshold for $y\rightarrow x$. Note that the existence of
a minimum and maximum coupling strengths for stable synchronization
in the last two cases is an unexpected occurrence for periodic
systems. The parameter values are $\sigma_r=0.1$, $\sigma_i=0.2$,
$l_r=0.5$ and $l_i=0.25$.}
\end{figure}
\begin{figure}[p]
 \centering
\includegraphics[width=0.85\textwidth]{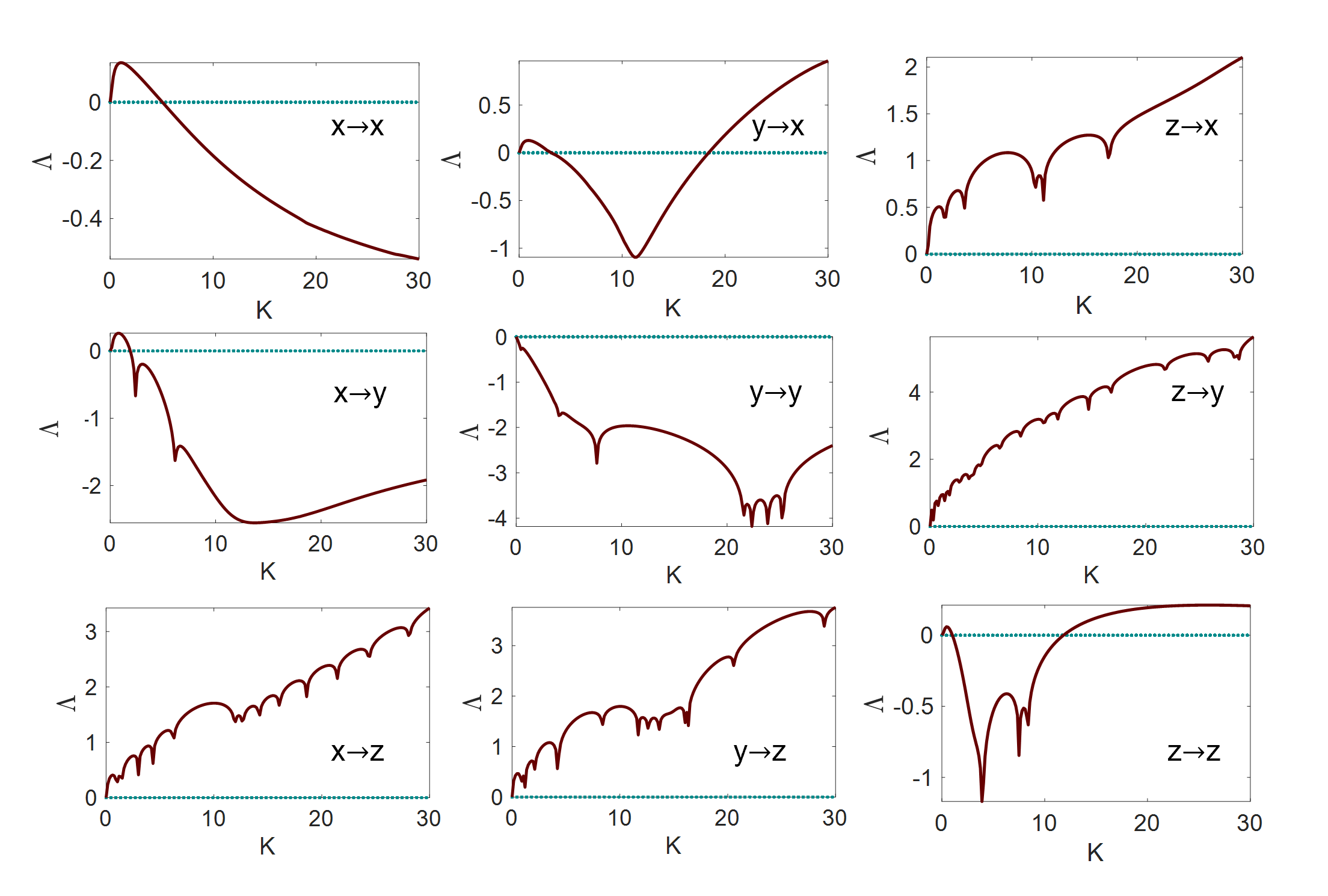}
\caption{\label{FigLorPer}\textbf{New classes
of synchronizability for the Lorenz system.} The Master Stability
Function ($\Lambda$) of the Lorenz system, defined in Eq.~\eqref{LorEq},
plotted as a function of the generalized coupling strength~$K$,
shows that it has four synchronizability classes. The MSF is always
positive for $z\rightarrow x$, $z\rightarrow y$, $x\rightarrow z$
and $x\rightarrow y$ couplings, it is negative after a threshold
for $x\rightarrow x$ and $x\rightarrow y$, it is negative within
a bounded range of coupling strengths for $y\rightarrow x$ and
$z\rightarrow z$, and it is always negative for $y\rightarrow y$.
Note that all cases except $y\rightarrow y$ challenge the current
beliefs about periodic systems, according to which one would expect
the MSF to be always negative. The parameter values are $\sigma=10$,
$\rho=28$ and $\beta=0.77$.}
\end{figure}

\section{The cabbage system}
The Cabbage system is a megastable, periodically-forced
oscillator with spatially-periodic damping~\cite{Spr17}.
This system has an infinite number of coexisting attractors,
and its dynamics is described by the system
\begin{equation}\label{cabEq}
 \begin{aligned}
  \dot x &= y\\
  \dot y &= -x + y\cos(x)\:.
 \end{aligned}
\end{equation}
Studying its Master Stability Function reveals that
the system is always in Class~II, corresponding to
an always stable synchronous state, except for the
coupling $y\rightarrow x$, for which there are two distinct, bounded regions of stability (Fig.~\ref{FigCabPer}).

\section{The Stuart-Landau oscillator}
The Stuart–Landau oscillator~\cite{Lan44,Stu60} is a well-studied
system because of its fundamental relevance to Hopf bifurcations,
and it has seen extensive use in the modelling of flow systems in
which supercritical bifurcations occur when a control parameter exceeds
a threshold. The system is defined by the equations
\begin{equation}\label{SLEq}
 \begin{aligned}
  \dot x &= \sigma_r x - \sigma_i y - \left( l_r x - l_i y\right) \left( x^2+y^2\right)\\
  \dot y &= \sigma_i x + \sigma_r y - \left( l_i x + l_r y\right) \left( x^2+y^2\right)\:.
 \end{aligned}
\end{equation}
Using parameter values $\sigma_r=0.1$, $\sigma_i=0.2$,
$l_r=0.5$ and $l_i=0.25$, we obtain a MSF that can belong
to three different classes. Specifically, it is in Class~II
for self-couplings, in Class~III for $x\rightarrow y$
and in Class~IV for $y\rightarrow x$ (Fig.~\ref{FigSLPer}).

\section{The Lorenz system}
The Lorenz system was first proposed for modeling and analyzing
the seemingly unpredictable behaviour of weather~\cite{Lor63}. Later,
it found wide applications in modeling different systems, including
lasers, batteries, and economic processes. The governing equations
of the Lorenz system are
\begin{equation}\label{LorEq}
 \begin{aligned}
  \dot x &= \sigma \left( y - x\right)\\
  \dot y &= x \left( \rho - z\right) - y\\
  \dot z &= xy - \beta z\:.
 \end{aligned}
\end{equation}
Choosing $\sigma=10$, $\rho=28$ and $\beta=0.77$,
the behaviour of the MSF identifies four possible
different classes for the system. More in detail,
for $z\rightarrow x$, $z\rightarrow y$, $x\rightarrow z$
and $x\rightarrow y$ couplings, the system is in
Class~I, and synchronization is never stable. For
$x\rightarrow x$ and $x\rightarrow y$, the system
is in Class~IV, with stable synchronization occurring
only after a threshold of coupling strength. For
$y\rightarrow x$ and $z\rightarrow z$, the system
is in Class~V, featuring a bounded region in which
the synchronous state is stable. Finally, for $y\rightarrow y$,
the MSF is always negative, and synchronization is
always stable (Fig.~\ref{FigLorPer}).